# UNA VERSIONE INTUIZIONISTA
# DEL TEOREMA DI RAMSEY PER COPPIE

Stefano Berardi
Universita' di Torino
Dipartimento di Informatica
http://www.di.unito.it/~stefano/

Torino, 10 gennaio 2014

RIASSUNTO. Il teorema di Ramsey [6] per coppie e' provabile classicamente ma non intuizionisticamente [2]. In questa nota mostriamo che Ramsey si puo' rienunciare in una forma intuizionisticamente provabile, informativa (o quanto meno priva di negazioni), e classicamente equivalente all'originale. A differenza di precedenti tentativi, non utilizziamo ne' il non-controesempio come in [1], [5], ne' aggiungiamo un nuovo principio all'intuizionismo come in [4]. Congetturiamo che questa <<versione intuizionista di Ramsey>> possa servire a rimpiazzare il Teorema di Ramsey nella prove di convergenza di programmi in [3].

INTRODUZIONE.
In questa nota mostriamo che Ramsey puo' essere rienunciato in una forma intuizionisticamente provabile e priva di negazioni, come segue:

*(H-chiusura per unioni finite)* Le relazioni binarie H-ben fondate sono chiuse per unioni finite:
$(R_1, \ldots, R_n \text{ H-ben fondate}) \Rightarrow (R_1 \cup \ldots \cup R_n \text{ H-ben fondata})$

Informalmente, una relazione binaria e' H-ben fondata se $H(R)$ e' ben fondato. $H(R)$ e' l'insieme delle sequenze $x_0, x_1, x_2, x_3, \ldots$ R-decrescente e transitive, cioe' tali che $i<j \Rightarrow x_j R x_i$ per ogni indice $i,j$. Rimandiamo a dopo una precisa formulazione intuizionista della buona fondazione. Per ora ci limitiamo a osservare che, classicamente, $H(R)$ ben fondato significa che non esistono sequenze infinite $x_0, x_1, x_2, x_3, \ldots$ tali che $i<j \Rightarrow x_j R x_i$ per ogni $i,j$ in Nat.

Vedremo che H-chiusura e' classicamente equivalente a Ramsey, con una prova molto piu' breve della piu' corta prova nota per Ramsey. In altre parole, la prova di Ramsey si puo' decomporre in una parte intuizionista all'incirca della lunghezza di una prova usuale di Ramsey, e che prova un enunciato abbastanza simile, seguita da una parte molto piu' breve ma classica.

La prova della H-chiusura e' divisa in tre capitoli. In § 0 enunciamo formalmente la H-chiusura. In § 1 diamo la nozione di ben fondato nell'intuizionismo. In § 2 proviamo sempre intuizionisticamente le proprieta' di base della buona fondazione. In § 3 diamo la prova vera e propria della H-chiusura.

Terminata la prova della H-chiusura, in § 4 proviamo classicamente che Ramsey e la H-chiusura si equivalgono.

---
§ 0 IL TEOREMA DI H-CHIUSURA
---

Usiamo le lettere I, J, … per indicare, R, S, T, U relazioni binarie, X, Y, Z sottoinsiemi, x, y, z, t, … elementi. Identifichiamo i sottoinsiemi X di I con le proprieta' degli elementi di I.

La definizione classica di R-ben fondata e': non esistono R-catene decrescenti infinite. Come definizione intuizionista di buona fondazione prendiamo una basata sulla nozione di proprieta' induttiva. Sia R una relazione binaria su I. Una proprieta' X e' R-induttiva se P passa da ogni yRx ad x stesso, un elemento x e' R-ben fondato se appartiene a ogni proprieta' induttiva, R e' ben fondato se ogni x in I e' R'ben fondato. Proviamo a scrivere le formule che esprimono "induttivo" e "ben fondato".

DEFINIZIONE 1 (buona fondazione intuizionista)
Sia R una relazione binaria su I e X un sottoinsieme di I
```
  IH^R_X(y)    = PerOgni z.(zRy => Xz)(ip. ind. per R in y per X)
  Ind^R_X(y)   = IH^R_X(y) => Xy        (X e' R-induttivo in y)
  IND^R_X      = PerOgni y.Ind^R_X(y)   (X e' R-induttivo)
  Wf^R_X(x)    = IND^R_X => Xx          (x e' R-ben fondato per X)
  WF^R(x)      = PerOgni X.Wf_X(x)      (x e' R-ben fondato)
  WF(R)        = PerOgni x.WF^R(x)      (R e' ben fondata)
```
*

Chiamiamo una coppia (I,R) con R relazione binaria su I una *struttura*, e diciamo che *(I,R) e' ben fondata* se R e' ben fondata.

Indichiamo coppie di un x in I e un y in J con (x,y). Denotiamo le liste su un insieme I con <x_1, …, x_n>: <> e' la lista vuota, <x> la lista del solo x eccetera. La concatenazione (.*.) tra liste sullo stesso insieme I e' definita come segue:
    <x_1, …, x_n> * < y_1, …, y_m> = <x_1, …, x_n, y_1, …, y_m>
Definiamo la relazione di estensione in un passo >_1 tra liste sullo stesso insieme I come: L*<y> >_1 L per ogni lista L.

Definiamo l'insieme H(T) delle successioni T-decrescenti transitive finite su I. H(T) e' l'insieme delle liste (anche vuote) <x_1, …, x_n> su I tali che:

Per ogni i,j in [1,n]: i<j => (x_j T x_i)
Ovvero: $x_2\ T\ x_1$; $x_3\ T\ x_2$, $x_3\ T\ x_1$; $x_4\ T\ x_3$, $x_4\ T\ x_2$, $x_4\ T\ x_1$, eccetera. Diciamo che T e' H-ben fondata se H(T) e' ben fondato per $>_1$ (per la relazione di estensione in un passo). Vogliamo provare il Teorema di Ramsey nella forma seguente:

TEOREMA DI H-CHIUSURA. Le relazioni binarie H-ben fondate su un insieme I sono chiuse per unioni finite.

---

## § 1 LA NOZIONE DI BUONA FONDAZIONE NELL'INTUIZIONISMO

---

Iniziamo introducendo delle definizioni che servono in piu' o meno ogni prova intuizionista di buona fondazione. Includiamo solo esempi e non prove.

L'esempio fondamentale di ben fondato e' Nat, < (i naturali con il loro ordine stretto). (*Prova di Nat,< ben fondato*. Dobbiamo provare che se X e' induttiva su Nat allora X = Nat. Proveremo che per ogni x in Nat abbiamo [0,x] incluso in X. Usiamo l'induzione di Peano per X: (X0 e PerOgni x in Nat.Xx=>X(x+1)) => (PerOgni z in Nat.Xz). X0 vale perche' non esistono x<0, dunque X vale per ogni x<0 e quindi, dato che X e' induttiva, vale per 0 e dunque in [0,0]. Supponiamo che X valga in [0,x]. Allora X vale in ogni y<x+1 in Nat, e quindi, dato che X e' induttiva, vale per x+1 e dunque in [0,x+1].)

Invece Z- = {0, -1, -2, …} non e' ben fondato per la relazione di predecessore $<_1$, in nessun punto. (*Prova di Z-,< non ben fondato*. < e' induttivo in ogni z Z- per la proprieta' X=vuoto, banalmente, perche' l'ipotesi induttiva: PerOgni z.(z $<_1$ x => Xz) ha controesempio z=x-1, dunque e' sempre falsa. Pero' la proprieta' X=vuoto e' sempre falsa, nonostante che sia una proprieta' induttiva. Ne segue che $<_1$ non e' ben fondato per nessun z in Z-).

Piu' in generale proveremo intuizionisticamente che se esiste una catena R-decrescente infinita da x:
   … $R\ x_2\ R\ x_1\ R\ x_0 = x$
allora R non e' ben fondato in x. Usando la logica classica e la scelta possiamo caratterizzare R ben fondato in x in termini di R-catene infinite: R e' ben fondato in x se e solo se <u>non</u> esistono catene R-decrescenti infinita da x. Dunque R e' ben fondato se e solo se <u>non</u> esistono catene R-decrescenti infinite in I. Questo risultato, equiparando la definizione intuizionista a quella classica, motiva la scelta della definizione intuizionista, anche se non puo' essere utilizzato in una prova intuizionista.

Diciamo che "R e' ben fondato verso l'alto" se il duale di R e' ben fondato, cioe' se la buona fondazione tratta di R-catene ascendenti anziche' discendenti. Per esempio, Nat, < non e' ben fondato verso l'alto, perche' esistono catene infinite per il >, che sono "decrescenti" nella nostra terminologia, e che possiamo scrivere cosi':

   … > 3 > 2 > 1 > 0

I criteri piu' usati per provare intuizionisticamente che una relazione e' ben fondata sono:
- un sottoinsieme di una relazione ben fondata e' ben fondata,
- se esiste un morfismo da una relazione R a una relazione S e S e' ben fondata lo e' anche R,
- se esiste una relazione di "simulazione" da R in S, allora ogni punto "simulabile" in un punto ben fondato e' a sua volta ben fondato. Cominciamo a introdurre la relativa terminologia, poi proveremo le relative proprieta' nella prossima sezione.

DEFINIZIONE 2 (Simulazioni).
Siano R, S due relazioni binarie, rispettivamente, su I, J. Sia T una relazione binaria su IxJ. Chiamiamo dominio di T l'insieme dom(T) = {x in I| Esiste y in J. xTy}.

 (a. morfismo) f:(I,R)->(J,S) e' un morfismo se f e' una funzione :I-> J tale che xRy => f(x)Sf(y), per ogni x,y in I.
 (b. Simulazioni) T e' una simulazione di R in S se:
    PerOgni x,y,z.( (xTy/\zRx) => Esiste t.(tSy/\zTt) )
 Una simulazione e' totale se dom(T) = I.
 (c. Simulabile) R e' simulabile in S se esiste una simulazione <u>totale</u> T di R in S.
 *

Possiamo descrivere il comportamento di una simulazione T di R in S mediante il completamento nell'angolo inferiore destro del seguente diagramma.

```
          T
   x -------------> y
   |                |
   |R               |S
   |        T       |
   z -------------> t
  insieme I     insieme J
```

Se abbiamo una simulazione e xTy possiamo trasformare ogni R-catena decrescente <u>finita</u> in I da x in una R-catena decrescente in J da y. Procediamo per completamento nell'angolo inferiore destro, aggiungendo in quest'ordine: y', y'', eccetera:

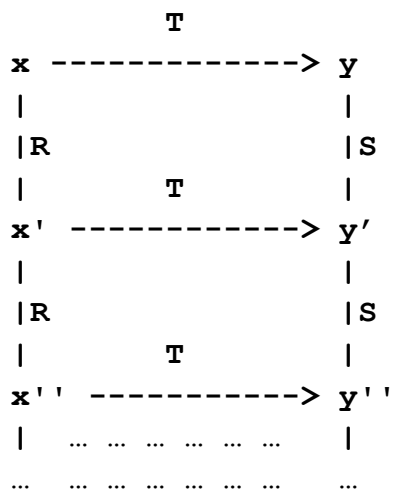

```
            T
    x ------------> y
    |               |
    |R              |S
    |       T       |
    x' -----------> y'
    |               |
    |R              |S
    |       T       |
    x'' ----------> y''
    |  … … … … … …  |
    … … … … … … …   …
```

Usando l'assioma della scelta questo risultato vale anche per le R-catene decrescenti <u>infinite</u> da un punto di dom(T). Dunque se non ci sono S-catene decrescenti infinite in J non ci sono R-catene decrescenti infinite in dom(T). Se, inoltre, la simulazione e' totale, non esistono R-catene decrescenti infinite neppure in I. Usando la logica classica e la scelta concludiamo che se S e' ben fondato e T e' totale anche R e' ben fondato. Piu' avanti vedremo una prova intuizionista dello stesso risultato (che non solo non usa la logica classica, ma neppure usa la scelta).

Qualche esempio di simulazione.
- Se esiste un morfismo f:(I,R) -> (J,S) allora R e' simulabile in S, con la simulazione totale T = {(x,f(x))| x in I}. Infatti se zRx allora f(z)Sf(x) per la definizione di morfismo, dunque possiamo completare il diagramma scegliendo t = f(z).
- Se R e' incluso in S allora R e' simulabile in S con la simulazione totale T = {(x,x) | x in I}. In questo caso completiamo il diagramma scegliendo t = y = x.

Esistono operazioni che producono relazioni ben fondate a partire da relazioni ben fondate. La piu' semplice e' l'operazione di "successore", che consiste nell'aggiunta di un "massimo". Se R e' una relazione su I, definiamo la relazione R+1 su I+1 come segue. Sia dato un elemento top <u>non in I</u>: poniamo
  I+1 = I u {top}
  R+1 = R u {(x,top) | x in I}
Definiamo la struttura "successore" di (I,R)) mediante (I,R)+1 = (I+1, R+1).

Un'altra operazione che definisce strutture ben fondate a partire da srutture ben fondate e' la struttura definita per componenti, ispirata all'ordine per componenti.

DEFINIZIONE 3 (La relazione R⊗S).
Siano R, S due relazioni, rispettivamente su I e su J. La relazione R⊗S di componenti R, S e' R⊗S =
   (R x Diagonale(J)) u (Diagonale(I) x S) u (RxS)
*

Alternativamente, R⊗S e' definita, per ogni x, x' in I e ogni y, y' in J, da: (x,y) R⊗S (x',y') <=>
   ((xRx')/\(y=y')) \/ ((x=x')/\(ySy')) \/ ((xRx')/\(ySy'))
Nel caso R,S siano ordini allora R⊗S e' l' ordine per componenti, detto anche ordine prodotto: in questo caso R⊗S = RxS. In generale, invece, R⊗S e' piu' grande di RxS.

Data una relazione R possiamo definire l'insieme degli alberi binari in cui il nodo figlio e' in relazione R con il nodo padre. Questo insieme risultera' essere ben fondato rispetto alla relazione di estensione in un passo tra alberi, purche' R sia ben fondata. La definizione di albero binario ha un ruolo molto importante in informatica teorica, e richiede una cura particolare. Gli alberi n-ari non sono semplici grafi finiti a forma di albero, ma hanno anche un punto speciale detto radice, al piu' due figli per ogni nodo, e un ordine tra i nodi figli di uno stesso padre.

Un albero binario finito su I e' definito induttivamente come un albero nullo detto nil, oppure una tripla di un elemento di I e di 2 alberi, detti sottoalberi immediati: abbiamo Tr=nil oppure Tr = <x,Tr1,Tr2>. Dunque l'insieme degli alberi binari finiti su I e' la minima soluzione della seguente equazione:
        TB(I) = nil + (I x TB(I) x TB(I))
Sia Tr = <x,Tr1,Tr2>. Diciamo che Tr e' un albero di radice x. Se Tr1 = Tr2 = nil diciamo che Tr e' un albero-foglia, se Tr1 =/= nil e Tr2 = nil diciamo che Tr ha esattamente un figlio sinistro, se Tr1 = nil e Tr2 =/= nil diciamo che Tr ha esattamente un figlio destro, se Tr1, Tr2 =/= nil diciamo che Tr ha un figlio destro e uno sinistro.

L'universo |Tr| di un albero binario su I e' l'insieme degli elementi di I presenti in Tr. Alternativamente, definiamo |Tr| per induzione su T:

|nil| = insieme vuoto
|<x,Tr_1,Tr_2>| = {x} u |Tr_1| u |Tr_2|

Se L = <x_1, …, x_n> e' una lista su I, definiamo l'universo di L come |L| = {x_1, …, x_n}. Diciamo che una lista L su I e' <u>coperta</u> da un albero binario Tr su I se |L| = |Tr|. La relazione di copertura ci servira' per simulare un insieme di liste in un insieme di alberi, in modo tale che una lista sia associata a un albero avente gli stessi elementi di I.

Un albero binario finito su I si puo' equivalentemente definire in molti altri modi. Per esempio, come un grafo orientato etichettato su I, vuoto (nel caso di nil) oppure con un elemento detto radice, con esattamente un cammino dalla radice a qualsiasi altro nodo, con i lati etichettati su C={1,2}, in modo tale che ogni etichetta compaia al piu' una volta nei lati che escono da ogni nodo. Chiamiamo C={1,2} l'insieme dei <u>colori</u>.

Alternativamente ancora, possiamo definire prima i rami colorati degli alberi binari, poi gli alberi binari come particolari insiemi di rami. Sia L = <x_1, …, x_n> una lista su I. Diciamo che L ha i <u>lati colorati</u> se L e' associata a una lista f = <c_1,…,c_{n-1}> su C={1,2}, con f=<> se L=<>. Immaginiamo che la lista L sia disegnata con il colore $c_i$ assegnato al segmento $x_i$, $x_{i+1}$, per ogni $1 \leq i \leq n-1$: in L ci sono quindi n-1 segmenti colorati. Tutti i punti della lista, tranne l'ultimo, sono seguiti da un segmento colorato.

Notiamo che, per definizione, (<>,<>) e' la lista colorata vuota, (<x>,<>) e' la lista colorata di un elemento: entrambe le liste, pur essendo "colorate" secondo la nostra definizione, non hanno lati e dunque non hanno in realta' colori.

Definiamo la relazione >_1 di estensione in un passo sulle liste con i lati colorati:
- (<x>, <>)         >_1 (<>,<>)
- (L*<y>, f*<c>)    >_1 (L,f) se L=/=<>

Se L=/=<>, diciamo che (L*<y>,f*<c>) e' una <u>estensione in un passo di colore c</u> di (L,f). f*<c> estende la colorazione f di L assegnando il colore c all'unico segmento nuovo di L*<y>, l'ultimo.

Possiamo ora definire equivalentemente un albero binario su I come un particolare insieme di liste colorate. Un albero binario Tr e' un insieme Tr di liste su I con lati colorati su {1,2}, tale che:
 a. (<>,<>) e' in Tr.
 b. c'e' al piu' un (<x>,<>) in Tr.
 c. Se (L*<y>,f*<c>) e' in Tr allora (L,f) e' in Tr.
 d. Siano: L=/=<>, (L,f) in Tr, c in C={1,2}. Allora esiste <u>al piu' una estensione in un passo</u> di (L,f) in Tr di colore c.

 Per esempio l'albero nullo consiste della sola lista (<>,<>): dunque nil = {(<>,<>)}. L'albero-foglia di radice x e' uguale a {(<x>,<>), (<>,<>)}. L'albero di figli y, z e' uguale a:
    {(<x,y>,<1>), (<x,z>,<1>), (<x>,<>), (<>,<>)}
 Per ogni albero Tr, la lista colorata vuota (<>,<>) ha al piu' una estensione in Tr, la lista (<x>,<>), dove x e' la radice di Tr. Ogni lista colorata (L,f) in Tr con L=/=<> ha al piu' una estensione (L*<y>,f*<1>) di colore 1 in Tr, e al piu' una estensione (L*<z>,f*<2>) di colore 2 in Tr.

 Abbiamo bisogno di definire una relazione Tr' >_1 Tr di estensione in un passo tra alberi binari, che consiste nell'aggiunta di una foglia a Tr.

DEFINIZIONE 4 (Estensione in un passo di alberi binari).
1. Se Tr'={(<x>,<>), (<>,<>)} e Tr={(<>,<>)} allora Tr' >_1 Tr.
2. Sia Tr un albero binario e:
 - (L,f) una lista colorata in Tr con L=/=<>;
 - (L*<y>,f*<c>) una estensione di (L,f) di colore c.
Se (L,f) non ha estensioni in Tr di colore c, allora
    Tr' = (Tr u {(L*<y>,f*<c>)})  >_1  Tr

 Terminiamo introducendo la nozione di co-induttivo. Diciamo che R e' co-induttivo in y per X se R soddisfa l'implicazione inversa di: "essere induttiva in y per X". Ovvero se la proprieta' X si propaga dal punto y a tutti gli R-predecessori di y (essere induttivo richiede la propagazione opposta):

 Coind^R_X(y)  = (IH^R_X(y) <= Xy)   (R e' co-induttivo in y per X)

 Con la nozione di co-induttivo concludiamo le definizioni utili nelle prove di buona fondazione. Ora iniziamo la prova di Ramsey2.

------------------------------------------
§ 2 LE PROPRIETA' DI BASE DELLA BUONA FONDAZIONE
------------------------------------------

Nella prima parte della prova dimostriamo risultati che servono in genere nelle prove intuizioniste di buona fondazione, e non sono, in realta', specifici per Ramsey. Il prossimo Lemma elenca i metodi piu' comuni per provare intuizionisticamente che una relazione e' ben fondata.

LEMMA 1. (proprieta' di base della buona fondazione). Siano I, J insiemi, R, S relazioni binarie rispettivamente su I, J.
 1. Essere ben fondato e' una proprieta' induttiva e coinduttiva:
    x e' R-ben fondato <=> PerOgni yRx: y e' R-ben fondato
 2. Se R, S sono ben fondate allora R⊗S e' ben fondata.
 3. Se R e' T-simulabile in S, se xTy e y e' S-ben fondato, allora x e' R-ben fondato.
 4. Se R e' T-simulabile in S ed S e' ben fondata, allora ogni x in dom(T) e' R-ben fondato.
 5. Se R e' simulabile in S ed S e' ben fondata, allora R lo e'.
 6. Se f:(I,R) -> (J,S) e' un morfismo, e S e' ben fondata, allora R e' ben fondata.
 7. Se R e' inclusa in S e S e' ben fondata allora R e' ben fondata.
 *

PROVA DEL LEMMA 1.
 1a. *La buona fondazione e' una proprieta' induttiva.* La prova e' per espansione delle definizioni. Supponiamo x in I e per ogni y in I: yRx => y e' R-ben fondato. Vogliamo provare che x e' ben fondato, cioe' che per ogni X induttivo, x e' in X. Per ogni y tale che yRx, abbiamo y in X per y ben fondato, dunque x e' in X dato che X e' induttivo.
 1b. *La buona fondazione e' una proprieta' coinduttiva.* Basta provare che la proprieta' X = {x | PerOgni y in I. yRx => y e' R-ben fondato} e' induttiva: allora X varra' per ogni x ben fondato, e quindi se x e' ben fondato e yRx allora anche y e' ben fondato.
 Assumiamo x in I e y in X per ogni yRx, al fine di provare che x e' in X, ovvero che ogni yRx e' ben fondato. Per ipotesi, y e' in X, e dunque ogni zRy e' ben fondato. Per 1a segue che y e' ben fondato, come volevasi dimostrare.
 2. Siano R, S ben fondate su I, J. Proviamo che per ogni x in I, (x,y) e' R⊗S-ben fondata, per ogni y in J. Procediamo per induzione su x e R, e assumiamo *(ipotesi induttiva principale)* che per ogni x'Rx, R⊗S e' ben fondata su (x',y'), per ogni y' in J. Ora proviamo che R⊗S e' ben fondata su (x,y) per induzione su y e S. Assumiamo che *(ipotesi induttiva secondaria)* che R⊗S e' ben fondata su (x,y'), per ogni y'Sy. Per 1a, per provare che (x,y) e'

R⊗S-ben fondata dobbiamo provare che ogni (x',y')R⊗S(x,y) e' R⊗S-ben fondata. Per definizione di R⊗S, abbiamo che x'Rx oppure che x'=x e y'Sy. Nel primo caso usiamo l'ipotesi induttiva principale, nel secondo caso l'ipotesi induttiva secondaria.

3. *Se R ristretto a X e' T-simulabile in S, se xTy e y e' ben fondato in (J,S), allora x e' ben fondato in (I,R).* Procediamo per induzione su y e S. Dobbiamo quindi provare che Y = {y in J | Perogni x in I.(xTy => x R-ben fondato)} e' S-induttivo. Assumiamo che per ogni t in J, se tRy allora t e' in Y, al fine di provare che y e' in Y. Sia xRy: dobbiamo provare che x e' ben fondato. Per 1a, dobbiamo provare che per ogni z in I, se zRx allora z e' ben fondato. Per definizione di simulazione, esiste un t in J tale che zTt e tSy. Per ipotesi e tSy abbiamo t in Y. Per definizione di Y e zTt segue che z e' ben fondato, come volevasi dimostrare.

4. *Se R e' T-simulabile in S ed S e' ben fondata, allora ogni x in dom(T) e' R-ben fondato.* Se x e' in dom(T) allora xTy per qualche y in J. Y e' S-ben fondato dato che S e' ben fondata, dunque x e' R-ben fondato per il punto 3.

5. *Se R e' simulabile in S ed S e' ben fondata, allora R lo e'.* Per definizione di simulabile esiste una simulazione T di R in S tale che I = dom(T). Per il punto 4, ogni x in I e' R-ben fondato.

6. Usiamo il punto 5 e la simulazione totale T={(x,f(x))|x in I}.

7. Usiamo il punto 5 e la simulazione totale T={(x,x)|x in I}.

    *

Qualche esempio di prova intuizionista di buona fondazione. Sia R una relazione binaria su I e sia x in I. Diciamo che *x e' R-minimale* se yRx per nessun y. Se x e' R-minimale allora x e' ben fondato per il Lemma 1.1: banalmente, per ogni y in I, se yRx allora y e' ben fondato, dato che non ci sono yRx. Ne segue che la relazione R = vuota, che rende ogni elemento di I R-minimale, e' ben fondata su ogni x in ogni I.

Diciamo *x ha T-altezza n in Nat* se il massimo numero di lati di una R-catena decrescente da x e' n (cioe' se la catena ha n+1 punti, contando x stesso). Sempre per il Lemma 1.1, per induzione su n possiamo provare che per ogni n in Nat, ogni punto di R-altezza <=n e' ben fondato. (*Prova.* Caso n=0. Se x ha R-altezza <=0 allora yRx per nessun y, dunque x e' R-minimale e usiamo l'esempio precedente. Assumiamo (vero per n) per provare (vero per n+1). Sia x di R-altezza <=n+1. Se yRx allora y ha altezza <=n, dunque y e' R-ben fondato per ipotesi induttiva. Per il Lemma 1.1 x e' R-ben fondato).

Continuiamo a derivare risultati di buona fondazione. Ricordiamo che la struttura (I,R)+1 e' definita come (I+1, R+1) = (I u {top}, R u {(x,top)|x in I}) per qualche elemento top <u>non in I</u>. Un semplice corollario del Lemma 1 dice:

COROLLARIO 1 (La struttura successore e' ben fondata).
    (I,R) ben fondata => (I,R)+1 ben fondata

PROVA. T={(x,x) | x in I} e' una simulazione di (I,R)+1 in (I,R). Infatti se x e' in I e y(R+1)x, allora yRx per definizione di R+1. Dunque se (I,R) e' ben fondato, allora ogni x in I e' R-ben fondato, dunque ogni x in I = dom(T) e' (R+1)-ben fondato per il Lemma 1.4. top e' ben fondato in (I,R)+1 per il Lemma 1.1, dato che se x(R+1)top allora x e' in I e dunque ben fondato in (I,R)+1. Dato che ogni x in I+1 e' in I oppure e' top, ne segue che (I,R)+1 e' ben fondato.
  *

Riprendiamo l'osservazione che Z- = {0, -1, -2, ...} con la relazione <_1 di predecessore <u>non</u> e' ben fondato. Osserviamo ora che, se esiste una R-catena decrescente infinita da x in I: ... R x_2 R x_1 R x_0 = x, allora esiste un morfismo (quindi una simulazione) f:(Z-,<_1) -> (I,R) definito da f(z) = x_z. Per il Lemma 1.6, se x fosse ben fondato in (I,R), allora 0 sarebbe ben fondato in Z-,<_1, contraddizione. Dunque abbiamo provato in modo intuizionista: se esiste una R-catena infinita decrescente da x, allora x <u>non</u> e' R-fondato. In altre parole la definizione intuizionista di ben fondato implica intuizionisticamente quella classica, mentre l'implicazione inversa e' solo classica (qui non spieghiamo perche').

Possiamo ora fornire esempi di insiemi non ben fondati intuizionisticamente. Nessun elemento x di Real e' ben fondato per <, dato che esiste la catena infinita decrescente ... < x-3 < x-2 < x-1 < x. Notiamo che, invece, ogni n in Nat e' ben fondato per <: dunque la buona fondazione non si preserva <u>aggiungendo elementi</u>. La buona fondazione non si preserva neppure aggiungendo <u>relazioni tra elementi</u>: Real e' ben fondato rispetto alla relazione vuota, ma non rispetto alla relazione <, che include la relazione vuota.

Il fatto che l'esistenza di una R-catena decrescente infinita neghi la buona fondazione ci consente di caratterizzare le relazioni R ben fondate e H-ben fondate quando I, R sono finiti (nel senso che abbiamo un elenco dei loro elementi). Sia k<omega la cardinalita' di I. Chiamiamo R-ciclo da x una sequenza x_n R

x_{n-1} R x_{n-2} R … R x_0, con x_n = x = x_0 e n>0. Chiamiamo R-cappio *(R-loop)* un R-ciclo di lunghezza 1, ovvero, un x in I tale che xRx.

Siano I,R finiti. Allora R e' ben fondata se e sono se non esistono R-cicli. (*Prova*. Dato che I,R sono finiti, possiamo decidere se esistono R-cicli. Supponiamo non che non esistano. Allora ogni R-catena ha punti a 2 a 2 distinti, quindi ha al piu' k punti. Dunque ogni x in I ha altezza <=k-1, quindi ogni x e' ben fondato. Supponiamo invece che esista un R-ciclo da x. Allora esiste una R-catena decrescente infinita da x, ottenuta ripetendo all'infinito il ciclo, quindi x non e' ben fondato).

Siano I,R finiti. R e' H-ben fondata se e solo se non esistono R-cappi. (*Prova.* Dato che I,R sono finiti, possiamo decidere se esistono R-cappi. Se non esistono R-cappi, allora non esiste una R-catena decrescente transitiva <x_0, …, x_n> tale che x_0 = x_n e n>0, perche' in tal caso avremmo x_0 R x_n, dunque x_0 sarebbe un R-cappio. Quindi H(T) non ha cicli, dunque le liste in H(T) non contengono ripetizioni, e son al piu' quante le permutazioni di I. Quindi H(T), >_1 e' ben fondata perche' e' una struttura finita senza >_1-cicli. Supponiamo invece che esista un R-cappio x. Allora per ogni n in Nat, ogni lista fatta di n ripetizioni di x e' R-decrescente transitiva, dunque H(T) e' mal fondata).

Notiamo che siamo ora in grado di provare il Teorema di H-chiusura per relazioni <u>finite</u> T_1, …, T_n su un insieme I finito. Per quanto appena provato, infatti, T = (T_1 u … u T_n) e' H-ben fondato se e solo se non esistono T-cappi. Questo equivale a: non esistono T_i-cappi per nessun i in [1,n], ovvero a: ogni T_i e' H-ben fondato. Nel resto du questa nota proveremo lo stesso risultato per T_1, …, T_n, I <u>qualunque</u>.

Passiamo ora a un risultato non immediato, che si puo' interpretare come un risultato di buona fondazione: il Lemma di Koenig. Il Lemma di Koenig dice: "se tutti i rami di un albero binario sono finiti allora l'albero e' finito". Il Lemma di Koenig e' un risultato solo classico. Esiste un corrispondente intuizionista, che intuizionisticamente e' piu' debole dell'originale, con "ben fondato" al posto di "finito" nell'ipotesi. La versione intuizionista e': "ogni albero binario ben fondato e' finito". Se usiamo la definizione intuizionista di ben fondato possiamo dare una prova intuizionista (che pero' qui non ci interessa).

Siamo interessati invece ad una versione di Koenig per alberi annidati (alberi i cui nodi sono essi stessi alberi), che chiameremo 2-Koenig. Consideriamo un albero Tr i cui nodi sono alberi binari finiti, con l'estensione in un passo >_1 come relazione padre/figlio tra i nodi/albero di Tr. Classicamente possiamo dire: se l'unione degli alberi binari in ogni ramo di Tr e' un albero binario con solo rami finiti, allora ogni ramo di Tr e' finito.

2-Koenig ci interessa perche' nella prova della versione intuizionista di Ramsey useremo una versione intuizionista di 2-Koenig, con "ben fondato" al posto di "finito". Nella prova di 2-Koenig usiamo l'usuale relazione di prefisso per liste su I: L <= M <=> Esiste una lista L' su I.(L*L' = M).

LEMMA 2 (2-Koenig intuizionista).
Sia C={1,2}. Sia LC un insieme di liste su I a colori in C. Sia TB(LC) l'insieme degli alberi binari su I che hanno tutti i rami in LC. Sia >_1 la relazione di estensione in un passo (tra liste oppure alberi). Allora:

   (LC, >_1) ben fondata => (TB(LC), >_1) ben fondata
*

PROVA DI 2-KOENIG. Ordiniamo le liste colorate (L,f), (M,g) per componenti: (L,f)<=(M,g) <=> (L e' prefisso di M) ed (f e' prefisso di g). Usiamo lambda, lambda', lambda_1, lambda_2, mu, … come variabili su liste colorate (L,f) in LC.

Per ogni lambda in LC, definiamo T(lambda) come l'insieme degli alberi binari Tr in TB(LC) tali che lambda e' in Tr, e per ogni mu in Tr: mu, lambda sono confrontabili nell'ordine per componenti.

Per ogni lambda, lambda' in LC, definiamo T(lambda,lambda') come l'insieme degli alberi binari Tr in TB(LC) tali che lambda, lambda' sono in Tr, ed ogni mu in Tr: (mu, lambda sono confrontabili) oppure (mu, lambda' sono confrontabili).

Proviamo che ogni struttura (T(lambda),>_1) e' ben fondata, per induzione su lambda in LC. Quindi porremo lambda = (<>,<>), la lista colorata vuota: ne seguira' che TB(LC) = T(lambda), dato che lambda appartiene ad ogni albero binario ed e' <= di ogni lista colorata. Dunque ne seguira' che (TB(LC),>_1) e' ben fondata.

Sia Tr in T(lambda). Per il Lemma 1.1, dobbiamo provare che ogni Tr' >_1 Tr e' ben fondato in (T(lambda),>_1). Per definizione di albero binario, Tr' contiene esattamente una lista colorata lambda_1 >_1 lambda, oppure esattamente due liste colorate lambda_1, lambda_2 >_1 lambda. Altrimenti, dato che Tr' contiene solo liste confrontabili con lambda, Tr' conterrebbe solo prefissi di lambda, che e' una lista in Tr, contraddicendo Tr' >_1 Tr. A seconda dei casi Tr' e' in T(lambda_1) oppure Tr' e' in T(lambda_1,lambda_2). Proviamo che ogni elemento di entrambi gli insiemi e' ben fondato in (T(lambda),>_1): ne seguira' che T(lambda),>_1 e' ben fondato, come voluto.

(a) Proviamo che ogni Tr' in T(lambda_1,lambda_2) e' ben fondato in (T(lambda),>_1). Ogni Tr' in T(lambda_1,lambda_2) si decompone come Tr'_1 u Tr'_2, dove Tr'_i = l'insieme dei mu in Tr confrontabili con lambda_i per i = 1,2. Infatti ogni mu in Tr' e' confrontabile con lambda, quindi con qualche lambda_i >_1 lambda in Tr'. Per costruzione Tr'_i in T(lambda_i) per i = 1,2. Supponiamo che Tr' e Tr' si decompongano rispettivamente come Tr'_1 u Tr'_2 e come Tr''_1 u Tr''_2. Se Tr'' >_1 Tr', allora (Tr''_1 >_1 Tr'_1 e Tr''_2 = Tr'_2), oppure (Tr''_1 = Tr'_1 e Tr''_2 >_1 Tr'_2), a seconda se il nuovo punto di Tr'' e' >=lambda_1 oppure >=lambda_2. Non sono possibili altri casi, altrimenti il nuovo punto di Tr'', essendo confrontabile con lambda_1 oppure lambda_2, sarebbe < lambda_1, lambda_2, dunque <= lambda, e quindi sarebbe in Tr'. Dunque la scomposizione di Tr' definisce una simulazione di T(lambda), >_1 nella struttura (T(lambda_1)xT(lambda_2), >_1⊗>_1). Per ipotesi induttiva T(lambda_1), >_1 e T(lambda_2), >_1 sono ben fondate, per il Lemma 1.2 anche (T(lambda_1)xT(lambda_2), >_1⊗>_1) e' ben fondata. La simulazione ha dominio il sottoinsieme T(lambda_1, lambda_2) di T(lambda). Per il Lemma 1.4 ogni Tr' in T(lambda_1, lambda_2) e' ben fondato in T(lambda), >_1.

(b) Proviamo che ogni Tr' in T(lambda_1) e' ben fondato in T(lambda), >_1. Non basta osservare che T(lambda_1), >_1 e' ben fondato: non e' detto che un elemento di T(lambda_1) sia ben fondato anche in T(lambda), >_1. Procediamo per induzione su Tr' in T(lambda_1), e proviamo che ogni Tr'' >_1 Tr' e' ben fondato in T(lambda), >_1. Se Tr'' e' ancora in T(lambda_1) applichiamo l'ipotesi induttiva su Tr''. Altrimenti Tr'' e' in T(lambda_1, lambda_2) per qualche lambda_2 >_1 lambda, dunque Tr'' e' ben fondato in T(lambda), >_1 per il ragionamento precedente.
*

---
§ 3 UNA PROVA INTUIZIONISTA DEL TEOREMA DI H-CHIUSURA
---

Sia T = la relazione vuota su I. Allora H(T) non contiene liste di lunghezza >=2, dunque H(T)= {<>, <x> | x in I}. H(T) e' ben

fondata per $>_1$, dato che ogni $<x>$ e' minimale, e $<>$ ha altezza $<=1$ (ha altezza 1 se I ha elementi). Dunque la relazione vuota e' H-ben fondata: per provare la H-chiusura, resta da provare che le relazioni H-ben fondate sono chiuse per unione binaria.

Introduciamo un particolare insieme di liste colorate, le $T_1,T_2$-liste colorate, che proveremo essere ben fondato se $T_1,T_2$ sono H-ben fondate. Sia $(L,f)$ una lista con lati colorati. Diciamo che $(L,f)$ e' una $T_1,T_2$-liste colorata se ogni volta che un segmento $x_i, x_{i+1}$ di $(L,f)$ ha colore $k=1,2$, allora $x_i$ e' $T_i$-maggiore di tutti gli elementi della lista che lo seguono (e non solo di $x_{i+1}$). Formalmente:

DEFINIZIONE 5 ($T_1,T_2$-liste colorate).
$(L,f)$ e' una $T_1,T_2$-lista colorata se $L=<x_1, …, x_n>$, $f=<c_1, …, c_{n-1}>$ (con $f=<>$ se $L=<>$) e se per ogni $1<=i<=n-1$ abbiamo:
$c_i = k$ => (per ogni $j$ in $[1,n]$: $i<j$ => ($x_j T_k x_i$))
$H(T_1,T_2)$ e' l'insieme delle $T_1,T_2$-liste colorate.
*

Chiamiamo $T_1,T_2$-albero un albero binario i cui rami (considerati come liste a 2 colori) sono tutti in $H(T_1,T_2)$. Indichiamo con $TB(T_1,T_2)$ l'insieme dei $T_1,T_2$-alberi. Consideriamo su $H(T_1,T_2)$ la relazione $>_1$ di estensione in un passo tra liste colorate, e su $TB(T_1,T_2)$ la relazione $>_1$ di estensione in un passo tra alberi binari.

Proviamo ora alcune proprieta' specifiche dell'H-buona fondazione. Ricordiamo che una lista su I e' coperta da un albero binario $Tr$ su I se $|L| = |Tr|$ (se gli elementi di L sono esattamente le etichette di $Tr$).

Lemma 3 (Simulazioni). Siano $T_1$, $T_2$ relazioni binarie su un insieme I.
1. $H(T_1,T_2)$, $>_1$ simulabile in $(H(T_1) \times H(T_2), >_1 \otimes >_1)+1$
2. $H(T_1 \cup T_2), >_1$ simulabile in $TB(T_1,T_2), >_1$

PROVA DEL LEMMA SIMULAZIONI.
1. Sia $(L,f) = (<x_1,…,x_n>, <c_1,…,c_{n-1}>)$ una lista colorata in $H(T_1,T_2)$, con $f=<>$ se $L=<>$. Definiamo una relazione T su $H(T_1,T_2) \times (H(T_1) \times H(T_2)+1)$ per casi.
 - Se $n=0$ poniamo: $T((L,f),top)$.
 - Se $n>0$ poniamo: $T((L,f),(L_1,L_2))$ se per ogni $i=1,2$: $L_i$ = la sottolista degli $x_j$ in $<x_1,…,x_n>$ tali che $c_j = i$.
Proviamo ora che T e' una simulazione totale.

*T e' una relazione totale* per costruzione.

*T e' una relazione* su $H(T\_1,T\_2) \times (H(T\_1) \times H(T\_2)+1)$. Se i e' in {1,2}, allora $L\_i$ e' in $H(T\_i)$. Infatti se $j<j'$ sono indici di $L\_i$, allora $c\_j=i$ e quindi $x\_{j'}$ T $x\_j$ per definizione di $T\_1,T\_2$-lista colorata (L,f).

*T e' una simulazione*. Distinguiamo due casi, a seconda se n=0 oppure n>0. Indichiamo con theta un elemento generico di $H(T\_1) \times H(T\_2)+1$.

*Sia n=0, quindi L=<>*. Supponiamo (L',f') >\_1 (L,f) e T((L,f),theta): allora theta=top, dunque T((L',f'),(L'\_1,L'\_2)) per definizione di T e perche' L' ha almeno un elemento. Infine (L'\_1, L'\_2) (>\_1 + 1) top per definizione di (>\_1 + 1_.

*Sia n>0, quindi L=/=<>*. Supponiamo (L',f') >\_1 (L,f) e T((L,f),theta). Per la definizione di >\_1 e L=/=<> abbiamo (L',f')=(L\*<y>,f\*<c>) per qualche y in I e qualche c in C={1,2}, e inoltre theta = (L\_1, L\_2), con L\_1, L\_2 definiti come sopra. Per qualche L'\_1, L'\_2 abbiamo T((L\*<y>,f\*<c>), (L'\_1,L'\_2)). Per la definizione di L'\_1, L'\_2, se c=1 abbiamo L'\_1 = L\_1\*<y> e L'\_2 = L\_2, se c=2 abbiamo L'\_1 = L\_1 e L'\_2 = L\_2\*<y>, e in entrambi i casi (L'\_1, L'\_2) >\_1⊗>\_1 (L\_1, L\_2), quindi anche (L'\_1, L'\_2) (>\_1⊗>\_1 + 1) (L\_1, L\_2), come volevasi dimostrare.

2. Come simulazione totale T(L,Tr) prendiamo la relazione |L| = |Tr| ("L e' coperta da Tr").

*T e' una simulazione totale*. Procediamo per induzione sulla lunghezza di L.

Abbiamo T(<>,nil) perche' |<>| = vuoto = |nil|.

Supponiamo che L = <x\_1,…,x\_n> e L\*<y> >\_1 L siano due liste ($T\_1$ u $T\_2$)-decrescenti transitive, e che |L| = |Tr| per qualche $T\_1,T\_2$-albero. Per definizione di simulazione, dobbiamo provare che per qualche Tr' >\_1 Tr abbiamo |L\*<y>| = |Tr'|. Per definizione di liste ($T\_1$ u $T\_2$)-decrescenti transitive, per ogni j in [1,n] abbiamo: y $T\_i$ $x\_j$ per qualche i in {1,2}. Per induzione su n, possiamo provare che esiste un h:[1,n]->{1,2} tale che, per ogni j in [1,n]: se i = h(j) allora y $T\_i$ $x\_j$. Vogliamo provare che esiste un $T\_1,T\_2$-albero Tr' >\_1 Tr in TB($T\_1$, $T\_2$), tale che |L\*<y>| = |Tr'|, ovvero tale che |Tr'| = |Tr| u {y}. Se Tr=nil prendiamo Tr'=l'albero-foglia di radice y. Supponiamo che Tr=/=nil, dunque n>=1, e sia $x\_1$ la radice di Tr. Definiamo un insieme J in cui scegliere la lista colorata da aggiungere a Tr. J consiste delle liste colorate della forma (M\*<y>,g) =
  ( <x\_{j\_1},…,x\_{j\_m}, y>, <h(x\_{j\_1}), …, h(x\_{j\_m})> )
per certi x\_{j\_1},…,x\_{j\_m} presi, eventualmente con ripetizioni, tra $x\_1$, …, $x\_n$, e tali che m>=1, e che la lista colorata (M,f) =
  ( <x\_{j\_1},…,x\_{j\_m}>, <h(x\_{j\_1}), …, h(x\_{j\_{m-1}})> )

sia in Tr.

*J non e' vuoto:* contiene almeno $(\langle x_1,y \rangle, \langle h(x_1) \rangle)$, con m=1.

*Ogni (M\*⟨y⟩,g) in J e' in H(T_1,T_2).* Infatti se (M,f) e' in Tr e dunque in H(T_1,T_2), e per ogni k in [1,m] se i = h(j_k) allora y $T_i$ x_{j_k} per la scelta di h.

Ci resta ora da provare il seguente Enunciato: <<*partendo da un qualunque (M\*⟨y⟩,g) in J possiamo trovare qualche (M'\*⟨y⟩,g') in J tale che Tr' = Tr u {(M\*⟨y⟩,g)} sia >_1 Tr e in TrBin(T_1,T_2)*>>. La prova e' per induzione inversa sulla lunghezza di (M,f) in Tr (che ha una lunghezza massima, dato che Tr e' finito).

*Prova dell'Enunciato.* Sia dato (M\*⟨y⟩,g) in J. (M\*⟨y⟩,g) e' in H(T_1,T_2) e (M,f) e' in Tr. Se (M,f) <u>non</u> ha una estensione di colore i = h(j_m) in Tr possiamo scegliere proprio (M\*⟨y⟩,g) per estendere Tr a un albero Tr' in TB(T_1,T_2). Altrimenti, (M,f) ha qualche estensione (M\*⟨x_{j_{m+1}}⟩, f\*⟨h(j_{m})⟩ ) in Tr. In questo caso J contiene una lista

   (M'\*⟨y⟩,g') = ( M\*⟨x_{j_{m+1}},y⟩, f\*⟨h(j_m),h(j_{m+1})⟩ ) associata a una lista (M',f') = (M\*⟨x_{j_{m+1}}⟩, f\*⟨h(j_{m})⟩) in Tr piu' lunga di (M,f). Per ipotesi induttiva otteniamo la tesi.
*

COROLLARIO 1. Siano T_1, T_2 relazioni binarie H-ben fondate su un insieme I.
1. L'insieme H(T_1,T_2), >_1 delle T_1,T_2-liste e' ben fondata.
2. L'insieme TrBin(T_1,T_2), >_1 dei T_1,T_2-alberi e' ben fondato.

PROVA. Siano H(T_1),>_1 e H(T_2),>_1 ben fondate.
1. H(T_1)xH(T_2), >_1⊗>_1) e' ben fondato per il Lemma 1.2, perche' le sue componenti lo sono. Per il Corollario 1, anche (H(T_1)xH(T_2),(>_1 ⊗ >_1))+1 e' ben fondato. H(T_1,T_2), >_1 e' simulabile in (H(T_1)xH(T_2),(>_1 ⊗ >_1))+1 per il Lemma 3.1, dunque e' ben fondato per il Lemma 1.5.
2. TB(T_1,T_2),>_1 e' ben fondato per il Teorema di 2-Koenig intuizionista, dato che H(T_1,T_2), >_1 e' ben fondata per il punto precedente.
*

TEOREMA DI H-CHIUSURA (versione intuizionista del Teorema di Ramsey).
Se T_1, …, T_n sono H-ben fondate, allora (T_1 u … u T_n) e' H-ben fondata.
*

PROVA (intuizionista). Per il Corollario 1.2, TrBin(T_1,T_2),>_1 e' ben fondato. Dunque le (T_1 u T_2)-catene decrescenti transitive, simulabili in TrBin(T_1,T_2),>_1 per il Lemma 3.2, sono ben fondate per il Lemma 1.5.
 *

COROLLARIO 2. Siano T_1, …, T_n relazioni binarie su un insieme I.
    T_1, …, T_n H-ben fondate <=> (T_1 u … u T_n) H-ben fondata

PROVA. L'implicazione => e' il Teorema precedente. Per provare l'implicazione <=, dato che ogni T_i e' incluso in (T_1 u … u T_n), basta provare che se S e' una relazione inclusa in T, allora T H-ben fondata implica S H-ben fondata. H(S) e' incluso in H(T). Per il Lemma 1.7, se H(T),>_1 e' ben fondato allora H(S),>_1 e' ben fondato. Dunque S inclusa in T e T H-ben fondata implica S H-ben fondata, come volevasi dimostrare.
 *

----------------------------------------------------------------
§ 4 UNA BREVE PROVA CLASSICA DELL'EQUIVALENZA TRA H-CHIUSURA E IL TEOREMA DI RAMSEY
----------------------------------------------------------------

Il Teorema di Ramsey per coppie dice: siano S, T due relazioni simmetriche e non riflessive su Nat, tali che (S U T) = {(x,y) in NxN|x=/=y}. Allora esiste un insieme infinito X incluso in Nat e <<omogeneo>>, ovvero tale che:

 (xSy per ogni x=/=y in X) \/ (xTy per ogni x=/=y in X)

Diamo ora una prova classica, che usa la scelta, dell'equivalenza tra H-chiusura e Ramsey.

 1. H-CHIUSURA=>RAMSEY
Assumiamo la H-chiusura. Siano S, T due relazioni simmetriche e non riflessive su Nat, tali che (S U T) = {(x,y) in NxN|x=/=y}: dobbiamo provare che esiste un insieme infinito X incluso in Nat e omogeneo. Siano S*={(x,y)|xSy /\ (x<y)} e T*={(x,y)|xTy /\ (x<y)}. Allora (S U T) = {(x,y)|x<y}, dunque (S U T) e' mal fondata e transitiva, quindi 0, 1, 2, 3, … e' una sequenza (S U T)-decrescente transitiva. Prendendo il contrappositivo della H-chiusura, esiste una catena infinita X={x_0, x_1, x_2, ...} in Nat per H(S*) oppure per H(T*). Nel primo caso: per ogni i<j in Nat abbiamo x_j S* x_i, dunque (x_i S x_j) e (x_j S x_i) per la simmetria di S. Nel secondo caso: per ogni i<j in Nat abbiamo x_j T* x_i, dunque (x_i T x_j) e (x_j T x_i) per la simmetria di T.

## 2. RAMSEY=>H-CHIUSURA

Assumiamo Ramsey, e supponiamo che esista una sequenza decrescente transitiva infinita $C = \{x_n \mid n \in \text{Nat}\}$ per (S U T), al fine di provare che ne esiste una per S oppure una per T. Dato che (S U T) e' transitiva abbiamo $i<j \Rightarrow x_j$ (S U T) $x_i$. Se prendiamo la chiusura simmetrica G di (S U T) otteniamo un grafo completo su Nat. Coloriamo di bianco un segmento (i,j) se (max(i,j) S min(i,j)), altrimenti di nero (e in questo caso abbiamo (max(i,j) T min(i,j)) dato che $i<j \Rightarrow x_j$ (S U T) $x_i$). Per Ramsey, esiste un insieme infinito $X = \{x_{n_0}, x_{n_1}, x_{n_2}, \ldots\}$ incluso in Nat con tutti i segmenti bianchi oppure tutti neri. Nel primo caso abbiamo $(i<j) \Rightarrow (x_{n_j} \ S \ x_{n_i})$ per ogni i,j in Nat, nel secondo $(i<j) \Rightarrow (x_{n_j} \ T \ x_{n_i})$ per ogni i,j in Nat. Dunque H(S) e' mal fondata con la sequenza S-decrescente transitiva infinita X, oppure H(T) e' mal fondata, con la stessa sequenza.